\documentclass[twocolumn,superscriptaddress,floatfix,preprintnumbers,nofootinbib]{revtex4-2}

\usepackage[colorlinks=true,breaklinks=true]{hyperref}
\usepackage[utf8]{inputenc}
\usepackage{longtable}
\usepackage{graphicx}
\usepackage{amsmath}
\usepackage{amssymb}
\usepackage{float}

\hypersetup{urlcolor=Blue, citecolor=Blue, linkcolor=Blue}
\usepackage{url}
\usepackage[normalem]{ulem}
\usepackage{mathtools}
\usepackage[dvipsnames]{xcolor}

\usepackage{orcidlink}

\newcommand{\nue}{\ensuremath{\nu_{e}}}
\newcommand{\nuebar}{\ensuremath{\bar{\nu}_{e}}}
\newcommand{\nux}{\ensuremath{\nu_{x}}}

\begin{document}

\title{Energy-dependent slow collective flavor conversion of supernova neutrinos}

\author{Heng-Hao Chen\orcidlink{0009-0006-8513-1473}} 
\email{henghaogapp@gmail.com}
\affiliation{Institute of Physics, Academia Sinica, Taipei 115201, Taiwan}
\affiliation{Institut f\"ur Astroteilchenphysik, Karlsruhe Institute of Technology (KIT), P.O. Box 3640, 76021 Karlsruhe,
Germany}

\author{Ian Padilla-Gay~\orcidlink{0000-0003-2472-3863}\,}
\email{ianpaga@berkeley.edu}
\affiliation{Department of Physics, University of California Berkeley, Berkeley, CA 94720}
\affiliation{Department of Physics, University of California San Diego, La Jolla, CA 92093}

\author{Meng-Ru Wu\orcidlink{0000-0003-4960-8706}}
\email{mwu@as.edu.tw}
\affiliation{Institute of Physics, Academia Sinica, Taipei 115201, Taiwan}
\affiliation{Institute of Astronomy and Astrophysics, Academia Sinica, Taipei 106319, Taiwan}
\affiliation{Physics Division, National Center for Theoretical Sciences, Taipei 106319, Taiwan}

\author{Sajad Abbar\orcidlink{0000-0001-8276-997X}}
\email{abbar@mpp.mpg.de}
\affiliation{Max-Planck-Institut {f\"ur} Physik (Werner-Heisenberg-Institut), Boltzmannstr. 8, 85748 Garching, Germany}

\author{Zewei Xiong\orcidlink{0000-0002-2385-6771}}
\email{z.xiong@gsi.de}
\affiliation{GSI Helmholtzzentrum {f\"ur} Schwerionenforschung, Planckstra{\ss}e 1, D-64291 Darmstadt, Germany}

\date{\today}

\preprint{N3AS-26-008}

\begin{abstract}
We study collective slow flavor conversion (SFC) of supernova neutrinos with multi-energy, multi-angle simulations for three representative neutrino spectra in the early accretion, late accretion, and cooling phases, in which multiple crossings between the initial electron- and heavy-lepton-flavor spectra are present. 
By numerically solving the neutrino quantum kinetic equations in a local periodic box, we find that SFC triggered predominantly by the spatially inhomogeneous instabilities drives the system toward a spatially coarse-grained, quasi-stationary state, whose flavor conversion probability depends strongly on energy, angle, and the neutrino mass ordering. 
While we find that not all of the initial spectral crossings are completely erased in the final state, a simple, box-like analytical prescription inspired by studies of fast flavor conversions, which eliminates the spectral crossings, can reasonably approximate the post-SFC spectra. 
Using the initial and post-SFC spectra, we also evaluate the changes of the corresponding $\nu_e$ and $\bar\nu_e$ heating rates as well as the absorption equilibrium electron fraction ($Y_e$). 
Within the considered scenarios, we find that the heating rates are generally enhanced by up to $\sim 80\%$ due to the net conversion of $\nu_x$ to $\nu_e$ (and $\bar\nu_x$ to $\bar\nu_e$) above their crossing energy, provided that the energy spectra above the crossing energy differ substantially.
For the absorption equilibrium $Y_e$, spectra changes due to SFC increase it by $\sim 0.03$ due to the relatively more enhanced $\nu_e$ absorption rate than $\bar\nu_e$, which potentially drives supernova materials to be more proton-rich.  
These results highlight the importance of energy-dependent treatments of SFC for supernova neutrinos.
\end{abstract}

\maketitle

\section{Introduction}
\label{sec:intro}

Dense neutrino gases in core-collapse supernovae (CCSNe) and binary neutron-star mergers can undergo collective flavor transformations driven by neutrino--neutrino coherent forward scattering~\cite{duan2010collective,volpe2024neutrinos,johns2025neutrino}. 
Among the different manifestations of collective oscillations, slow flavor conversion (SFC) arises from the interplay between vacuum oscillations and neutrino self-interactions~\cite{duan2006collective,hannestad2006selfinduced}. 
By modifying the electron-flavor energy spectra, SFC can affect neutrino heating, the electron fraction $Y_e$ of the ejecta, and the resulting heavy-element nucleosynthesis~\cite{duan2011influence,dasgupta2012role,wu2015effects}. 
Due to the recently much-improved understanding of collective flavor oscillations, SFC has received renewed attention in studies of its instability properties~\cite{fiorillo2025theory1,fiorillo2025theory,dasgupta2025sufficient}, spatiotemporal evolution~\cite{padillagay2025flavor}\footnote{We note that, during the completion of this manuscript, Ref.~\cite{goimilgarcia2026quasisteady}
which performed a similar study for SFC has appeared. 
Our work presented here has been done independently using a different
simulation code. While the results of the two studies broadly agree, quantitative differences regarding the coarse-grained asymptotic state exist, which motivate  different analytical approximations (see Secs.~\ref{sec:results} and \ref{sec:recipe}).}, and its relevance in realistic astrophysical settings~\cite{shalgar2024neutrino,fiorillo2025neutrinomassdriven,froustey2026neutrino}. 

In Ref.~\cite{padillagay2025flavor}, we examined SFC in a single-energy, multi-angle system and demonstrated that its asymptotic state depends sensitively on the \nuebar/\nue~number-density ratio and more modestly on the neutrino mass ordering. 
A single-energy treatment, however, cannot describe the different mean energies and spectral shapes of \nue, \nuebar, and \nux, nor the resulting spectral crossings. 
These features allow different energy modes to experience different amounts of flavor conversion and are particularly important for charged-current rates of neutrinos with nucleons, whose energy weighting makes them sensitive to the high-energy tails of the post-SFC spectra.

In this work, we extend our previous study to a multi-energy, multi-angle system for three representative CCSN phases: early accretion, late accretion, and cooling. 
For both normal and inverted mass orderings, we perform the linear stability analysis and solve the nonlinear quantum kinetic equations in a one-dimensional periodic box until the system has reached a quasi-stationary state after being coarse-grained over the simulation volume. 
We then discuss the effect of SFC on neutrino energy spectra, quantify the changes in the \nue~and \nuebar~absorption heating rates and in equilibrium $Y_e$, and test a phenomenological box prescription that estimates the asymptotic spectra directly from the initial lepton-number spectrum. 

We find that SFC is predominantly triggered by spatially inhomogeneous instabilities.  
The coarse-grained, quasistationary flavor conversion probability in the final state depends on the energy, angle, and neutrino mass ordering. 
Interestingly, unlike in the single-energy system studied in Ref.~\cite{padillagay2025flavor}, the initial spectral crossings are not completely erased in the coarse-grained, angle-averaged final state.
Nevertheless, it can still be reasonably approximated by an analytical prescription inspired by studies for fast flavor conversions~\cite{xiong2023evaluating,zaizen2023simple}. 
We also evaluate the corresponding changes of the $\nu_e$ and $\bar\nu_e$ heating rates as well as the associated equilibrium electron fraction ($Y_e)$ due to SFC, finding appreciable deviations from their initial (unoscillated) values. 

The paper is organized as follows. We introduce in Sec.~\ref{sec:init_spec} the initial neutrino energy spectra  considered in this work, and describe the quantum kinetic equations, initial conditions, and numerical setup in 
Sec.~\ref{sec:qke_setup}. 
Linear stability analysis and nonlinear evolution are presented in Sec.~\ref{sec:lsa} and Sec.~\ref{sec:results}, respectively. 
Sec.~\ref{sec:astro_implications} quantifies the impact of SFC on the physical energy spectra, the resulting heating rates, and the equilibrium $Y_e$. 
In Sec.~\ref{sec:recipe}, we introduce the phenomenological box prescription and compare it with the simulation outcome.
Sec.~\ref{sec:conclusion} summarizes our conclusions. 
In Appendix~\ref{appendix:resolution}, we discuss the numerical convergence and the required simulation resolution.
Appendix~\ref{app:eigenfit} includes the complete coarse-grained asymptotic phase-space distributions and compares them with the dominant unstable eigenfunctions. 
We use natural units with $\hbar=c=1$ throughout the paper.

\section{Neutrino energy spectra in benchmark scenarios}
\label{sec:init_spec}

\begin{figure*}[t]
\centering
\includegraphics[width=0.94\textwidth]{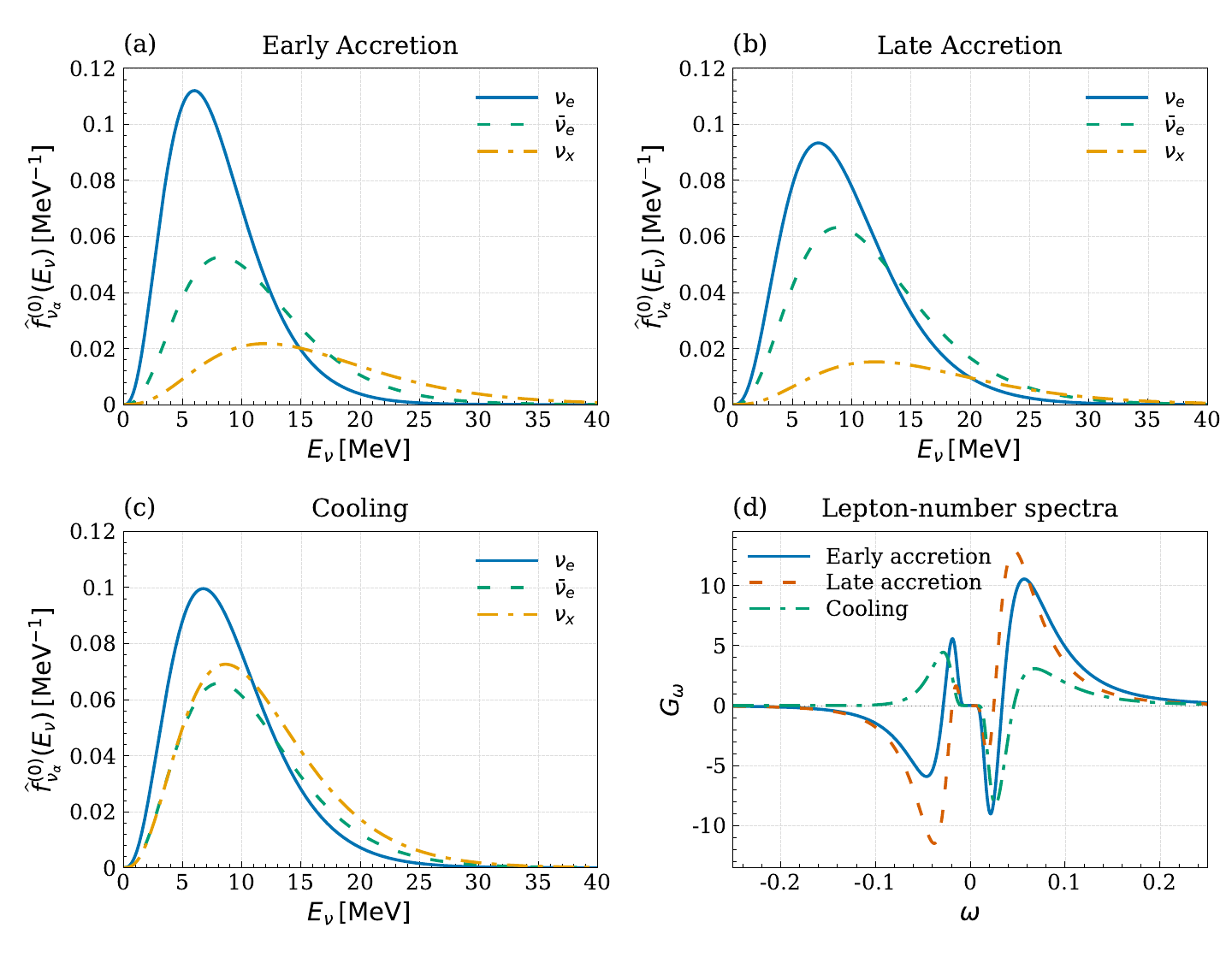}
\caption{Initial spectra of the three benchmark phases. 
Panels (a)--(c) show the $\nu_e$-normalized energy spectra $\widehat f_{\nu_\alpha}^{(0)}(E_\nu)$ from
Eq.~\eqref{eq:fhat_initial} using $\langle E_{\nu_\alpha}\rangle$ and $L_{\nu_\alpha}$ values listed in Table~\ref{tab:benchmarks} for early accretion, late accretion, and cooling, respectively. 
Panel (d) shows the corresponding lepton-number spectra $G_\omega$ [Eq.~\eqref{eq:Gomega}]. 
Note that panel (d) takes the dimensionless convention adopted throughout the text by setting $\mu=\sqrt{2}G_Fn_{\nu_e}^{(0)}=1$ so that $\omega$ is in units of $\mu$.
}
\label{fig:initial_benchmark_spectra}
\end{figure*}

We work in the two-flavor scenario that includes $\nu_e$, $\nu_x$, and their antiparticles, with $\nu_x$ denoting the heavy-lepton neutrino, $\nu_\mu$ or $\nu_\tau$, whose properties are taken to be identical. 
We take three sets of neutrino mean energies and energy luminosities for different species at different post-bounce times, listed in Table~\ref{tab:benchmarks}.
These values are taken from three snapshots at post-core-bounce time $t_{\rm pb}=0.05$, $0.2$, and $3.0$~s from a reference spherically symmetric CCSN simulation for an 18~$M_\odot$ progenitor~\cite{fischer2010protoneutron}, and represent typical conditions during the early accretion, late accretion, and cooling phases, respectively.

For each species $\nu_\alpha=\nu_e,\bar\nu_e,\nu_x,\bar\nu_x$, we define the number-density normalization factor relative to the initial $\nu_e$ as

\begin{equation}
\label{eq:Fnu}
 R_{\nu_\alpha}\equiv
\frac{n_{\nu_\alpha}^{(0)}}{n_{\nu_e}^{(0)}}=
 \frac{L_{\nu_\alpha}/\langle E_{\nu_\alpha}\rangle}
 {L_{\nu_e}/\langle E_{\nu_e}\rangle}\,,
\end{equation}
where $n_{\nu_\alpha}^{(0)}$ is the initial number density of $\nu_\alpha$ in the considered location. 
The initially $\nu_e$-normalized, angle-integrated energy spectra for a given species are assumed to follow the parametrization of Ref.~\cite{keil2002montecarlo}: 
\begin{align}
\widehat f_{\nu_\alpha}^{(0)}(E_\nu)
 &=R_{\nu_\alpha}
\frac{(\chi_{\nu_\alpha}+1)^{\chi_{\nu_\alpha}+1}}
{\Gamma(\chi_{\nu_\alpha}+1)}
\frac{E_\nu^{\chi_{\nu_\alpha}}}
{\langle E_{\nu_\alpha}\rangle^{\chi_{\nu_\alpha}+1}}\nonumber\\[-2pt]
&\quad\times
\exp\!\left[-(\chi_{\nu_\alpha}+1)
\frac{E_\nu}{\langle E_{\nu_\alpha}\rangle}\right],
\label{eq:fhat_initial}
\end{align}
where $E_\nu$ is the neutrino energy and $\Gamma(x)$ is the Gamma function.
Note that $\int_0^\infty dE_\nu\,\widehat f_{\nu_\alpha}^{(0)}(E_\nu)=R_{\nu_\alpha}$.
We set $\chi_{\nu_\alpha}=3$ for all species, so their differences are entirely specified by the mean energies and luminosities in Table~\ref{tab:benchmarks}.

\begin{table}[t]
\centering
\caption{Benchmark mean energies and energy luminosities for the three
CCSN phases taken at different post-bounce time $t_{\rm pb}$ from Ref.~\cite{fischer2010protoneutron}. 
Values in each entry separated by slashes are given in the order of $\nu_e$, $\bar\nu_e$, and $\nu_x$, respectively. 
We assume $\nu_x$ and $\bar\nu_x$ share identical properties.}
\label{tab:benchmarks}
\begin{tabular}{lccc}
\hline\hline
Phase & $t_{\rm pb}$ [s] & $\langle E_{\nu_\alpha}\rangle$ [MeV] & $L_{\nu_\alpha}$ [$10^{51}$~erg/s]\\
\hline
Early accretion & $0.05$ & $8 / 11 / 16$ & $36 / 32 / 28$ \\
Late accretion  & $0.2$  & $9.6 / 11.8 / 16$ & $44 / 45 / 20$ \\
Cooling         & $3.0$  & $9.0 / 10.8 / 11.5$ & $2.1 / 2.0 / 2.5$ \\
\hline\hline
\end{tabular}
\end{table}

Panels (a)--(c) of Fig.~\ref{fig:initial_benchmark_spectra} show the initial spectra for different species, assuming that $\nu_x$ and $\bar\nu_x$ are identical before flavor oscillations occur.
In each phase, the electron- and heavy-flavor spectra cross once in both the neutrino and antineutrino sectors. 
The $\nu_e$--$\nu_x$ and $\bar\nu_e$--$\bar\nu_x$ crossings occur at $E_\nu\simeq14.87$ and $17.66$~MeV for the early accretion phase spectra, $20.06$ and $26.22$~MeV for late accretion, and $10.88$ and $4.03$~MeV for cooling, respectively. 
At lower energies (below the crossing energy), the electron-flavor spectrum is larger, while the heavy-flavor spectrum dominates above it. 
For the cooling phase spectra, the low-energy excess of $\bar\nu_e$ over $\bar\nu_x$ below $4.03$~MeV is practically negligible.

\subsection{Lepton-number spectra}
\label{subsec:Spectrum_Gomega}

For SFC, it is useful to define the lepton-number spectra $G_\omega$ as a function of the vacuum oscillation frequency $\omega=\pm|\delta m^2|/(2E_\nu)$, where $+$ ($-$) corresponds to $\nu$ ($\bar\nu$) and $|\delta m^2|$ is the absolute value of the neutrino mass-squared difference.
For convenience, we work in dimensionless units by setting the representative neutrino-neutrino forward-scattering strength parameter $\mu\equiv \sqrt{2}G_F n_{\nu_e}^{(0)}=1$, with $G_F$ the Fermi coupling constant, and express all dimensional quantities in units of $\mu$ throughout this paper unless specified.
We set $\omega=0.05$ for $E_\nu=10$~MeV. 

Following Eq.~\eqref{eq:fhat_initial}, the normalized spectrum of each species in the $\omega$ space becomes
\begin{equation}
f_{\nu_\alpha}(\omega)
=\frac{[(\chi_{\nu_\alpha}+1)\bar\omega_{\nu_\alpha}]
^{\chi_{\nu_\alpha}+1}}
{\Gamma(\chi_{\nu_\alpha}+1)}
|\omega|^{-(\chi_{\nu_\alpha}+2)}
e^{-(\chi_{\nu_\alpha}+1)\bar\omega_{\nu_\alpha}/|\omega|},
\label{eq:gwapa}
\end{equation}
where $\bar\omega_{\nu_\alpha}
\equiv|\delta m^2|/(2\langle E_{\nu_\alpha}\rangle)$ so that $\int_0^\infty d\omega\,f_{\nu_\alpha}(\omega)=1$ for neutrinos and $\int_{-\infty}^0 d\omega\,f_{\bar\nu_\alpha}(\omega)=1$ for antineutrinos.

Figure~\ref{fig:initial_benchmark_spectra}(d) shows $G_\omega$ defined by 
\begin{equation}\label{eq:Gomega}
G_\omega\equiv 
\begin{cases}
\displaystyle G^+_\omega = f_{\nu_e}(\omega)
-R_{\nu_x}f_{\nu_x}(\omega),
&\omega>0,\\[9pt]
\displaystyle G^-_\omega =R_{\bar\nu_x}f_{\bar\nu_x}(\omega)
-R_{\bar\nu_e}f_{\bar\nu_e}(\omega),
&\omega<0, 
\end{cases}
\end{equation}
for all three benchmark phases. 
Positive $G_\omega$ at $\omega>0$ denotes a $\nu_e$ excess, whereas negative $G_\omega$ at $\omega<0$ denotes a $\bar\nu_e$ excess relative to the corresponding heavy flavor.
The zeros of $G_\omega$, which correspond to crossings between the electron- and heavy-lepton-flavor energy spectra, are located at $\omega\simeq-0.0283$ and $0.0336$ for the early accretion phase, $\omega\simeq-0.0191$ and $0.0249$ for late accretion, and $\omega\simeq-0.124$ and $0.0460$ for cooling. 
In all three phases, $G_\omega$ also changes sign across $\omega=0$, where the high-energy tails vanish.
While there exist three zero crossings in $G_\omega$ for all three phases, the spectrum for the cooling phase essentially represents a two-crossing case, since the antineutrino sector is practically dominated by $\bar\nu_x$.

\section{Quantum kinetic equations and numerical setup}
\label{sec:qke_setup}

We adopt the convention that treats both $\nu$ and $\bar\nu$ on equal footing~\cite{duan2006collective,dasgupta2009multiple}, and use $\rho(\omega>0)$ and $\rho(\omega<0)$ to denote their respective reduced density matrices. 
Assuming translation symmetry in the $x$ and $y$ directions and axisymmetry with respect to the local $z$ direction, the evolution of the dense neutrino gas without momentum-changing collisions is governed by the mean-field quantum kinetic equation (QKE),
\begin{eqnarray}\label{eq:qkenu}
    (\partial_t + v_z\partial_z)\,\rho(t,z,\omega,v_z) &=& -i\,[H_{\rm vac} + H_{\nu\nu},\, \rho] \ ,
\end{eqnarray}
where the diagonal and off-diagonal elements of 
\begin{eqnarray}\label{eq:rho}
    \rho = \begin{pmatrix} 
    \rho_{ee} & \rho_{ex} \\
    \rho_{ex}^{*} & \rho_{xx} 
    \end{pmatrix}
    \ ,
\end{eqnarray}
are related to the differential electron-minus-heavy-lepton number (E-XLN) per phase space volume (see the next subsection) and flavor correlations for both $\nu$ ($\omega>0$) and $\bar\nu$ ($\omega<0$), respectively.  
Note that we assume spatial homogeneity of the matter term originating from the neutrino-electron forward scattering and use an effective in-medium vacuum mixing angle to account for its effect~\cite{estebanpretel2008role}. 

The vacuum Hamiltonian in the above QKE is given by
\begin{equation}
H_{\rm vac}=\frac{\eta_{\rm MO}\omega}{2}
\begin{pmatrix}
-\cos2\theta_V&\sin2\theta_V\\
\sin2\theta_V&\cos2\theta_V
\end{pmatrix},
\label{eq:Hvac}
\end{equation}
where $\eta_{\rm MO}=+1$ ($-1$) for normal (inverted) mass ordering and $\theta_V\ll1$ is the effective in-medium mixing angle. 
The neutrino self-interaction Hamiltonian is
\begin{eqnarray}\label{eq:Hnunu}
H_{\nu\nu} &=& 
\int_{-\infty}^{\infty} d\omega' \int d v_z'\,(1 - v_z v_z') \rho(\omega', v_z') \ .
\end{eqnarray}
Since we have set $\mu=\sqrt{2}G_F n_{\nu_e}^{(0)}=1$, it does not explicitly appear in $H_{\nu\nu}$.

\subsection{Initial conditions and numerical setup}

We initialize the reduced density matrices by
\begin{align}
\rho(t_0,z,\omega,v_z)
&=\frac{G(\omega,v_z)}{2}
\begin{pmatrix}1+s(z)&\varepsilon(z)\\
\varepsilon(z)&1-s(z)\end{pmatrix},
\label{eq:rho_initial}
\end{align}
where
\begin{equation}\label{eq:g_omega_vz}
G(\omega,v_z)=G_\omega g_\nu(v_z)
\end{equation}
with $g_\nu(v_z)\propto \exp[-(v_z-1)^2/(2\sigma_\nu^2)]$, a normalized angular distribution that satisfies $\int dv_z\,g_\nu(v_z)=1$.  
Clearly, $G(\omega,v_z)$ represents the $\nu_e$-normalized differential E-XLN spectrum with $\int d\omega dv_z G(\omega,v_z)=1-R_{\nu_x}+R_{\bar\nu_x}-R_{\bar\nu_e}$ after integrating over the entire $\omega$ and $v_z$ ranges.
We adopt $\sigma_\nu=0.6$ for every species so that no E-XLN crossing is present in the initial condition, i.e., the flavor conversion studied here is driven by crossings in frequency space.
We take $\varepsilon(z)=10^{-2}\exp\!\left(-z^2/50\right)$ to seed inhomogeneous and real off-diagonal perturbations. 
Correspondingly, $s(z)=\sqrt{1-\varepsilon^2(z)}$ so that the reduced density matrices satisfy $\operatorname{Tr}\rho=G$.
The seed changes the initial diagonal spectra only at $\mathcal O(\varepsilon^2)$, which we neglect when quoting the initial spectra and number densities.

Given the initial conditions, we solve the QKE for each phase until the system has reached a spatially averaged (coarse-grained) quasi-stationary state at about $t\simeq 1000$, using the finite difference version of \texttt{COSE$\nu$}~\cite{george2023cosenu} extended for this work to support the energy dependence explicitly. 
The simulations use periodic boundary conditions over $z\in[-50, 50]$, a uniform velocity grid over $v_z\in[-1,1]$, and a uniform frequency grid over $\omega\in[-0.25, 0.25]$.
Unless stated otherwise, the results shown below are obtained with $N_z = 1000$, $N_{v_z} = 600$, $N_\omega = 200$, and the Courant--Friedrichs--Lewy (CFL) number of 0.9 for the two accretion phases; for the cooling phase, we use $N_z = 2000$, $N_{v_z} = 800$, and $N_\omega = 200$. 
The resolution dependence on some of these choices is provided in Appendix~\ref{appendix:resolution}.

Following Ref.~\cite{padillagay2025flavor}, we introduce the polarization vector $\mathbf{P}$ of each $(\omega,v_z)$ mode, whose components are given by
\begin{eqnarray}
 P_1 &\equiv& +2\,\mathrm{Re}(\rho_{ex})/G \, , \nonumber \\
 P_2 &\equiv& -2\,\mathrm{Im}(\rho_{ex})/G \,, \nonumber \\
 P_3 &\equiv& (\rho_{ee}-\rho_{xx})/G \, . 
\end{eqnarray}
Initially, $\mathbf P=(\varepsilon,0,s)$, and unitarity preserves the norm of each polarization vector. 
Equivalently, the evolved reduced density matrix can be written as
\begin{equation}
\rho=\frac{G}{2}
 \begin{pmatrix}1+P_3&P_1-iP_2\\P_1+iP_2&1-P_3\end{pmatrix} \ .
\end{equation}
The survival probabilities for an initial $\nu_e$, $\bar\nu_e$, $\nu_x$, or $\bar\nu_x$ can then be expressed in terms of $P_3$ as 
\begin{equation}
 P_{\rm sur}(\omega,v_z)=\frac{1+P_3(\omega,v_z)}{2} \ .
 \label{eq:survival_probabilities}
\end{equation}

With these definitions, the evolved neutrino distributions at any given time for all species can be written as 
\begin{align}
\Phi_{\nu_e}(\omega, v_z)&=R_{\nu_x}f_{\nu_x}(\omega)g_\nu(v_z)+G(\omega,v_z) P_{\rm sur}(\omega, v_z) \, , \nonumber\\
\Phi_{\nu_x}(\omega, v_z)&=R_{\nu_e}f_{\nu_e}(\omega)g_\nu(v_z)-G(\omega,v_z)P_{\rm sur}(\omega, v_z) \, ,\nonumber\\
\Phi_{\bar\nu_e}(\omega, v_z)&=R_{\bar\nu_x}f_{\bar\nu_x}(|\omega|)g_\nu(v_z)
-G(\omega,v_z) P_{\rm sur}(\omega, v_z) \, ,\nonumber\\
\Phi_{\bar\nu_x}(\omega, v_z)&=R_{\bar\nu_e}f_{\bar\nu_e}(|\omega|)g_\nu(v_z)
+ G(\omega,v_z) P_{\rm sur}(\omega, v_z) \, , 
\label{eq:Phi_reconstruction}
\end{align}
where $\Phi_{\nu_e}$ and $\Phi_{\nu_x}$ ($\Phi_{\bar\nu_e}$ and $\Phi_{\bar\nu_x}$) are defined for $\omega>0$ ($\omega<0$).

From Eq.~\eqref{eq:Phi_reconstruction}, one can compute the time-dependent, spatial- and angular-averaged neutrino energy spectra as 

\begin{equation}
 \begin{aligned}
 \widehat f_{\nu_\alpha}(E_\nu,t)
 &=\int dv_z\,
 \left\langle\Phi_{\nu_\alpha}
 \!\left(t,z,v_z,\omega(E_\nu)\right)\right\rangle_z
 \left|\frac{d\omega}{dE_\nu}\right|,
 \end{aligned}
 \label{eq:fhat_from_phi}
\end{equation}
where $\left|d\omega/ dE_\nu\right|=|\omega/E_\nu|$ is the Jacobian. 
One can show that Eq.~\eqref{eq:fhat_from_phi} reduces to Eq.~\eqref{eq:fhat_initial} at the initial time.

\section{Linear Stability Analysis}
\label{sec:lsa}

The linear stability analysis (LSA) helps determine whether a small perturbation in flavor correlation for a given system configuration will grow exponentially over time (signaling an unstable solution)~\cite{banerjee2011linearized,izaguirre2017fast}.  
It also helps determine the spatial wavenumber at which the instability develops, which constrains the length scales over which flavor evolution takes place. 
We develop the LSA adapted to our discretized multi-energy, multi-angle system.

\begin{figure}[ht]
\centering
\includegraphics[width=1.0\columnwidth]{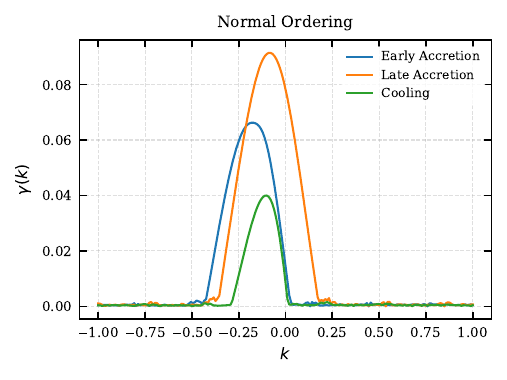}
\includegraphics[width=1.0\columnwidth]{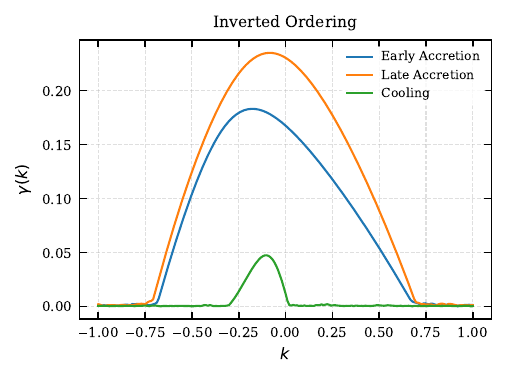}
\caption{Unstable growth rate $\mathrm{Im}\,\Omega=\gamma(k)$ for the three benchmark phases in normal
(upper) and inverted (lower) mass orderings. 
Table~\ref{tab:gamma_max} reports the maximum growth rate for each phase. 
Note that the two panels adopt different $y$ scale. The growth rates in IO are generally larger than in NO.
The values of $\gamma$ and $k$ are in units of $\mu$.
}
\label{fig:cool_dispersion}
\end{figure}

We linearize the reduced density matrices around the flavor-diagonal state following Ref.~\cite{padillagay2025flavor}, writing
\begin{equation}\label{eq:rho_lin}
    \rho(t,z,\omega,v_z)
    =\begin{pmatrix}
    \rho_{ee}^{(0)}-\Delta & G S/2\\
    G S^*/2 & \rho_{xx}^{(0)}+\Delta
    \end{pmatrix}.
\end{equation}
Here, $\rho_{\alpha\alpha}^{(0)}$ denotes the unperturbed reduced spectra obtained by taking $\varepsilon\to0$ in the initial conditions above. 
Thus, $\rho_{ee}^{(0)}=G$, $\rho_{xx}^{(0)}=0$ for both neutrinos and antineutrinos.
The perturbation is represented by $S=P_1-iP_2=2\rho_{ex}/G$.  
Note that unitarity gives $\Delta=G[1-\sqrt{1-|S|^2}]/2=\mathcal O(|S|^2)$ and therefore the diagonal evolution is subleading in the linear regime. 
As a result, only the equation for $S$ is retained.

\begin{table}[t]
\centering
\caption{Dominant growth rate $\gamma_\mathrm{max}$ extracted from Eq.~\eqref{eq:gamma_def}, and the wavenumber $k_\star$ at which it develops, for the three CCSN phases.
The values of $\gamma_{\rm max}$ and $k_*$ are in units of $\mu$.}
\label{tab:gamma_max}
\begin{tabular}{lcc}
\hline\hline
Phase & $\gamma_\mathrm{max}$ (NO/IO) & $k_\star$ \\
\hline
Early accretion & $0.066$/$0.18$ & $-0.18$\\
Late accretion  & $0.091$/$0.23$ & $-0.08$\\
Cooling         & $0.039$/$0.047$ & $-0.10$\\
\hline\hline
\end{tabular}
\end{table}

Substituting the commonly used plane-wave ansatz
\begin{equation}\label{eq:ansatz}
S(t,z,\omega,v_z)=Q(\omega,v_z)\,e^{-i(\Omega t+kz)},
\end{equation}
into Eq.~\eqref{eq:qkenu} and keeping only terms linear in $Q$, one obtains the following eigenvalue equation
\begin{equation}\label{eq:lsa}
    \Omega\,\vec Q = M(k)\,\vec Q\ ,
\end{equation}
where $\vec Q$ has components $Q_{ij} \equiv Q(\omega_i, v_j)$ on the discrete $(\omega, v_z)$ grid. 
The dominant (largest) growth rate at wavenumber $k$ is
\begin{equation}\label{eq:gamma_def}
    \gamma(k) \equiv \max_{n}\,\mathrm{Im}\,\Omega_n(k)\ ,
\end{equation}
where $\Omega_n(k)$ is the $n$th eigenvalue of the matrix $M(k)$ and $n$ labels the discrete eigenmodes. 
The overall fastest-growing mode of the system is $\gamma_\mathrm{max} \equiv \max_k\,\gamma(k)$, which corresponds to the maximum value among all $k$ modes.
The matrix elements of $M$ read
\begin{align}
M_{(i,a)(j,b)} =& \, \delta_{ij}\delta_{ab}\,\big[
-\eta_{\rm MO}\omega_i-kv_a
+\eta(1-v_a\langle v\rangle_g)\big]\nonumber \\
&-G_{\omega_j}g_\nu(v_b)(1-v_av_b)\,\Delta\omega\,\Delta v\ ,\label{eq:Mmatrix}
\end{align}
where the $(i,a)$ and $(j,b)$ indices label the $(\omega,v_z)$ pairs, and
\begin{equation}\label{eq:eta-vg}
\eta\equiv\int d\omega\,G_\omega\,,\qquad
\langle v\rangle_g\equiv\int dv_z\,v_z\,g_\nu(v_z)\,.
\end{equation}
These are the integrated lepton-number asymmetry and the velocity-weighted mean of the angular distribution. 
To avoid clutter in Eqs.~\eqref{eq:Mmatrix}, we have dropped the subscript $z$ on the velocity variables.

\begin{figure*}[ht]
\centering
\includegraphics[width= 1.0\textwidth]{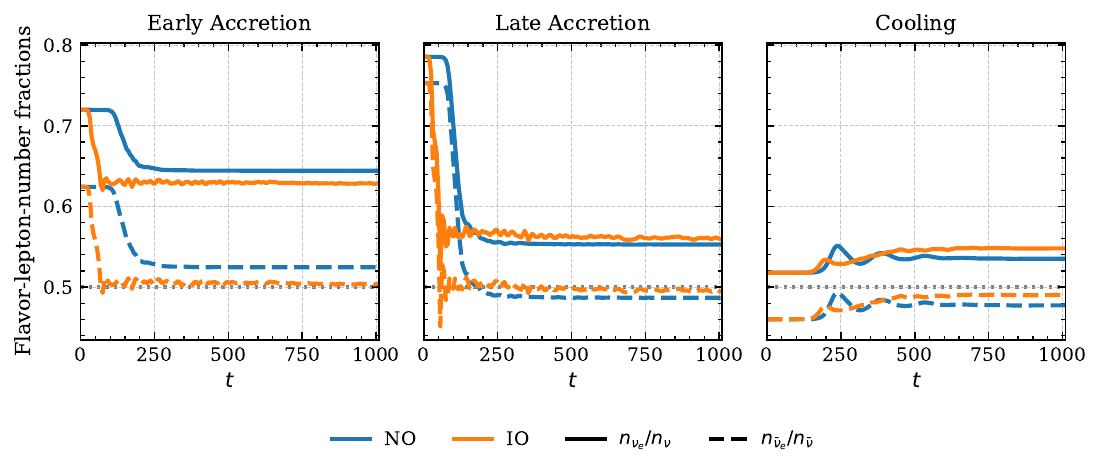}
\caption{Time evolution of the flavor-lepton-number fractions $n_{\nu_e}/n_\nu$ (solid) and $n_{\bar\nu_e}/n_{\bar\nu}$ (dashed) for the early accretion (left panel), late accretion (middle panel), and cooling (right panel) phases, in NO (blue) and IO (orange), with $t$ in units of $\mu^{-1}$. 
Here $n_\nu \equiv n_{\nu_e} + n_{\nu_x}$ and $n_{\bar\nu}\equiv n_{\bar\nu_e}+n_{\bar\nu_x}$, both energy- and angle-integrated and averaged over the box. 
The horizontal dotted line at $1/2$ marks flavor equipartition.
}
\label{fig:time_evolution}
\end{figure*}

For each supernova phase of Table~\ref{tab:benchmarks}, we construct $M(k)$ on a uniform velocity grid in $v_z\in[-1,1]$ with $N_v=100$ and a uniform grid of $N_\omega=80$ over $\omega\in[-0.25,0.25]$. 
The matrix is diagonalized with a standard dense eigenvalue solver from \texttt{numpy}~\cite{harris2020array} at each of $N_k=199$ wavenumbers uniformly distributed in $k\in[-1,1]$, and the dominant growth rate $\mathrm{Im}\,\Omega$ is kept for each $k$.
The dispersion relation $\gamma(k)$ identifies a band of unstable wavenumbers in each phase, with the maximum growth rate $\gamma_\mathrm{max}$ tabulated in Table~\ref{tab:gamma_max}. 
The wavenumber at which the maximum is reached, defined as $k_\star$, is non-zero in all three benchmarks, showing that the instability is inhomogeneous, in agreement with our previous findings in Ref.~\cite{padillagay2025flavor}.

Figure~\ref{fig:cool_dispersion} shows the dispersion relations for all three phases in both mass orderings. 
For the cooling phase in NO, the instability is confined to a narrow band of negative wavenumbers around $k \simeq -0.1$ and reaches a peak growth rate of $\gamma_\mathrm{max} \simeq 0.039$; modes outside this band have growth rates compatible with zero within numerical noise from spurious modes. 
The two accretion phases behave differently in the two orderings: in NO their unstable bands are only slightly broader than the cooling one, with $\gamma_\mathrm{max} \simeq 0.066$ and $0.091$ for early and late accretion, respectively, whereas in IO both bands extend over $|k| \lesssim 0.7$ and the peak growth rates increase to $0.18$ and $0.23$. The cooling phase, by contrast, is almost insensitive to the ordering ($0.039$ versus $0.047$).
This difference likely originates from the different shape of $G_\omega$ shown in Fig.~\ref{fig:initial_benchmark_spectra}(d).

Moreover, $k_\star$ can be cross-checked from the initial background (E-X)LN current,
\begin{align}
J_z&\equiv\int d\omega\,dv_z\,v_z\,G(\omega,v_z)
=\eta\langle v\rangle_g \ . 
\end{align}
Notice that the diagonal terms in Eq.~\eqref{eq:Mmatrix} combine as $-v_a(k+J_z)$. 
Empirically, we find that for all three phases and both mass orderings, the fastest-growing mode is the zero mode of the shifted wavenumber, $k'=k+J_z=0$, and therefore $k_\star=-J_z$, independently of the mass ordering. 
The shifted wavenumber absorbs the phase induced by the current $J_z$, following the dispersion relation formulation of Ref.~\cite{izaguirre2017fast}. In other words, when $k^\prime=0$ the diagonal term proportional to $v_a k^\prime$ vanishes in Eq.~\eqref{eq:Mmatrix}, allowing the instability to remain coherent. 
Numerically, this gives $k_\star=(-0.185,-0.088,-0.108)$ for the early-accretion, late-accretion, and cooling phases, respectively, in agreement with Table~\ref{tab:gamma_max}.

The LSA growth rates estimated here (Fig.~\ref{fig:cool_dispersion}) agree well with those extracted from the early exponential growth of the nonlinear solutions presented in the next section for all three supernova phases.

\begin{figure*}[ht]
\centering
\includegraphics[width= 1.0\textwidth]{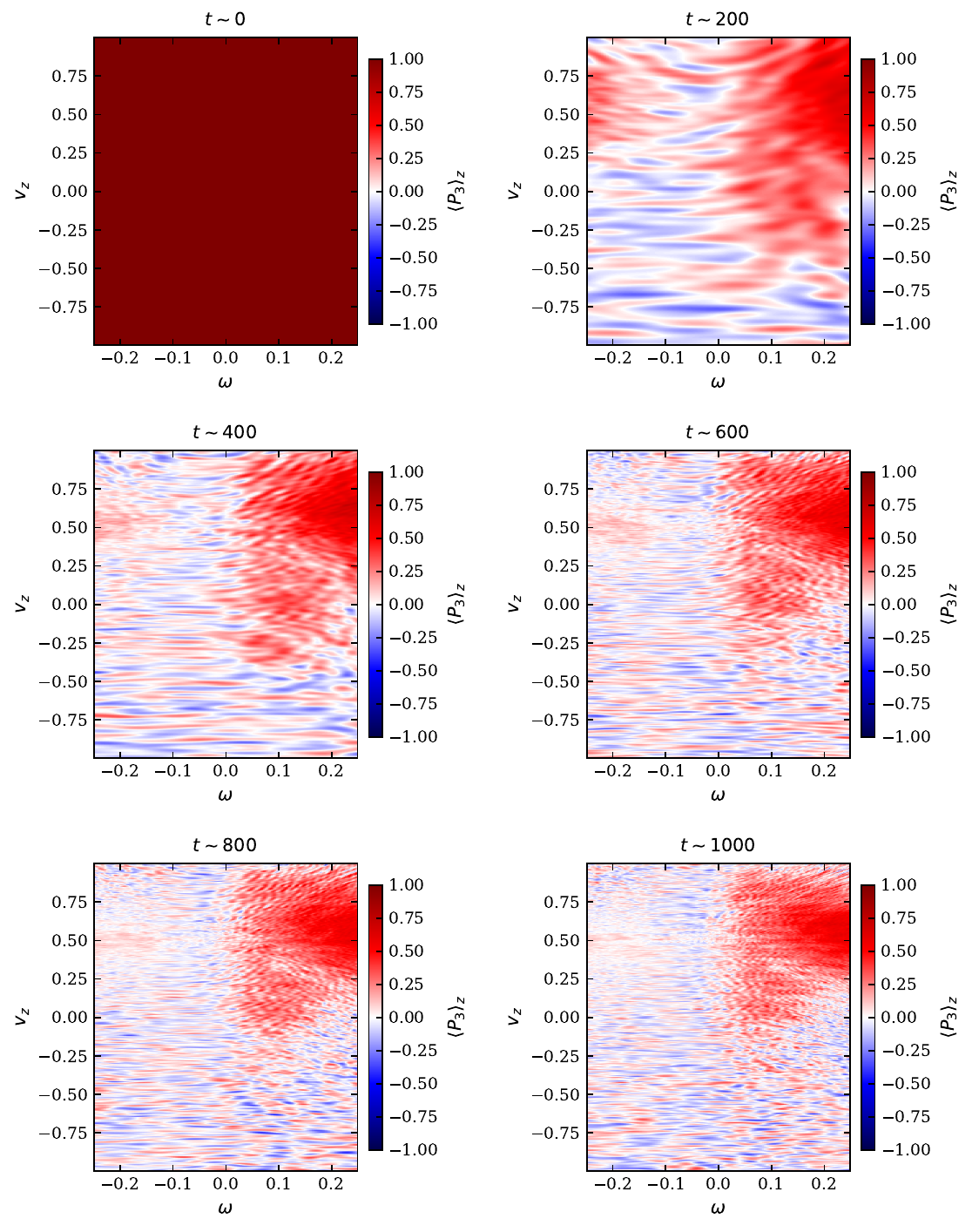}
\caption{Snapshots of the spatially averaged flavor polarization $\langle P_3\rangle_z$ in the $(\omega, v_z)$ plane at six times, for the late accretion phase in IO. 
The quantities $t$ and $\omega$ are in units of $\mu^{-1}$ and $\mu$, respectively. 
}
\label{fig:late_accr_evo}
\end{figure*}

\begin{figure*}[ht] 
\centering
\includegraphics[width= 0.9\textwidth]{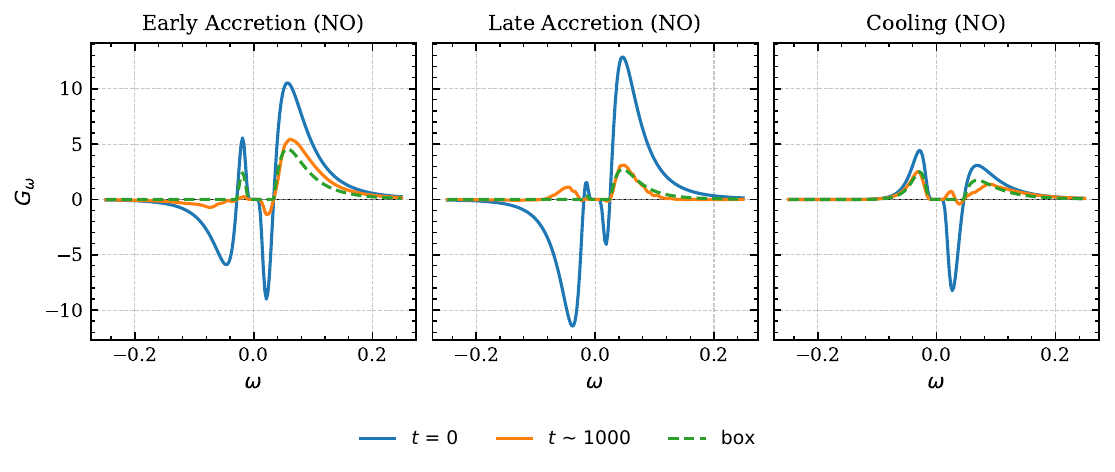}
\caption{Comparison of the initial ($t=0$, blue) and asymptotic ($t\simeq 1000$, orange) lepton-number spectra $G_\omega$, averaged over $z$ and integrated over $v_z$, for the early accretion (left panel), late accretion (middle panel), and cooling (right panel) phases in NO. 
Also shown are the final $G_\omega$ computed with the box recipe (green dashed curve); see Sec.~\ref{sec:recipe} for details. 
The quantities $t$ and $\omega$ are in units of $\mu^{-1}$ and $\mu$, respectively.
}
\label{fig:w_spec_NO}
\end{figure*}

\begin{figure*}[ht]
    \centering
    \includegraphics[width= 0.9\textwidth]{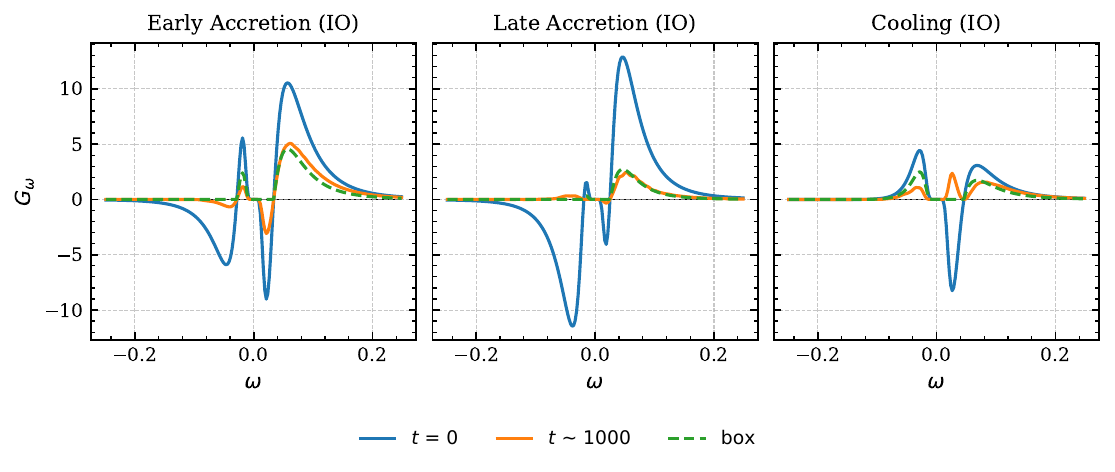}
    \caption{Same as Fig.~\ref{fig:w_spec_NO}, but for IO.}
    \label{fig:w_spec_IO}
\end{figure*}
\section{Non-linear Flavor Evolution}
\label{sec:results}

We show in Fig.~\ref{fig:time_evolution} the evolution of the coarse-grained,  energy- and angle-integrated fractions $n_{\nu_e}/n_\nu$ and $n_{\bar\nu_e}/n_{\bar\nu}$ for all three phases and both mass orderings, where $n_\nu \equiv n_{\nu_e} + n_{\nu_x}$ and $n_{\bar\nu}\equiv n_{\bar\nu_e}+n_{\bar\nu_x}$. 
In every case, the ratios stay near their initial values during the linear phase, depart rapidly once the instability saturates, and then evolve toward quasi-stationary values by $t\simeq1000$. 
The saturation time tracks the growth rates of Table~\ref{tab:gamma_max}: the late accretion phase, which has the largest $\gamma_\mathrm{max}$, departs from its initial state first, while the cooling phase is the slowest.

For the early and late accretion phases, both $n_{\nu_e}/n_\nu$ and $n_{\bar\nu_e}/n_{\bar\nu}$ decrease from their initial values to lower final values, reflecting the net conversion of electron flavor to heavy-lepton flavor because $R_{\nu_e}/R_{\nu_x}>1$ and $R_{\bar\nu_e}/R_{\bar\nu_x}>1$. 
In contrast, $R_{\nu_e}/R_{\nu_x}<1$ and $R_{\bar\nu_e}/R_{\bar\nu_x}<1$ for the cooling phase, so that SFC results in a net increase of $n_{\nu_e}/n_\nu$ and $n_{\bar\nu_e}/n_{\bar\nu}$.
For all phases, the different mass orderings only moderately affect the final values by $\lesssim 0.03$, consistent with the findings of Ref.~\cite{padillagay2025flavor}.
For all cases, $n_{\bar\nu_e}/n_{\bar\nu}$ in IO appears to reach a value closer to 0.5, i.e., equipartition, than in NO, which is also consistent with Ref.~\cite{padillagay2025flavor}.
For $n_{\bar\nu_e}/n_{\bar\nu}$, they asymptote to final values constrained by approximate E-XLN conservation~\cite{padillagay2025flavor}.

Fig.~\ref{fig:late_accr_evo} further shows the spatially averaged polarization vector component $\langle P_3\rangle_z$ in the $(\omega, v_z)$ plane at six times for the late accretion phase in IO as an example. 
First, $\langle P_3\rangle_z$ develops a pronounced dependence on both energy and angle. 
The neutrino modes with $\omega \gtrsim \omega_c \simeq 0.0249$ and with $v_z \gtrsim 0$ retain $\langle P_3\rangle_z \simeq 0.3$--$0.8$ and therefore undergo little flavor conversion. 
In contrast, the entire antineutrino half-axis, the high-energy neutrinos with $\omega \lesssim \omega_c$, and the backward-moving modes with $v_z \lesssim 0$ all relax to $\langle P_3\rangle_z \simeq 0$, i.e., near flavor equipartition. 
Second, both the angular and the energy structures become progressively finer with time. 
In particular, the cascade in $v_z$ is stronger, producing finer horizontal stripes; see Appendix~\ref{appendix:resolution} for further discussion.

\begin{figure*}[ht] 
\centering
\includegraphics[width= 1.0\textwidth]{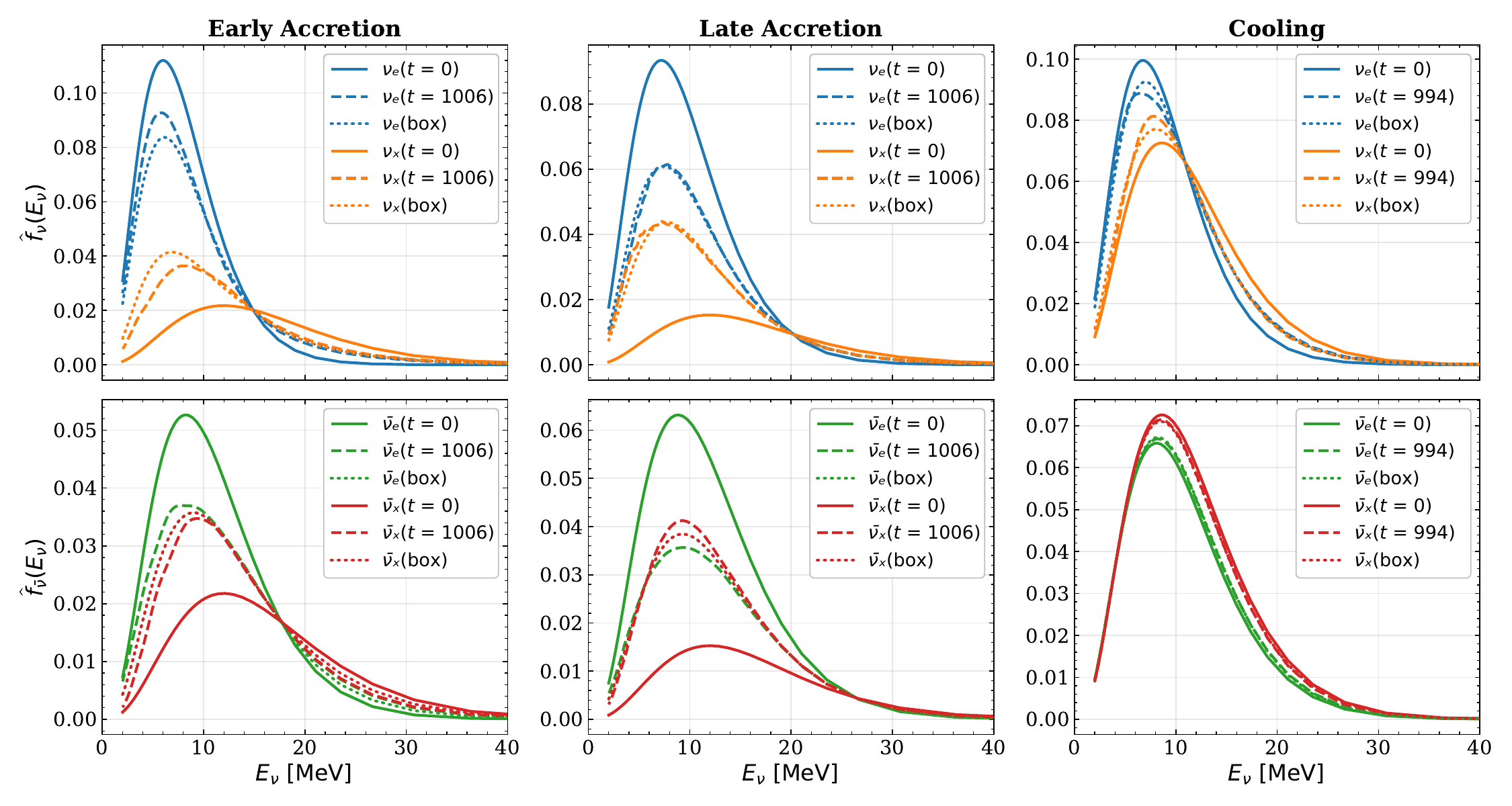}
\caption{Comparison of the initial ($t=0$, solid) and asymptotic ($t\simeq 1000$, dashed) energy spectra $\widehat f_{\nu_\alpha}(E_\nu)$, reconstructed with Eq.~\eqref{eq:fhat_from_phi}, for the early accretion (left panel), late accretion (middle panel), and cooling (right panel) phases in NO. 
The upper row shows $\nu_e$ (blue) and $\nu_x$ (orange), and the lower row shows $\bar\nu_e$ (green) and $\bar\nu_x$ (red). 
The spectra are averaged over space and integrated over velocity. 
The color-matched dotted curves show the spectra computed using the box recipe (see Sec.~\ref{sec:recipe}).
$t$ is in units of $\mu^{-1}$.
}
\label{fig:energy_spectra_no}
\end{figure*}

\begin{figure*}[ht]
\centering
\includegraphics[width= 1.0\textwidth]{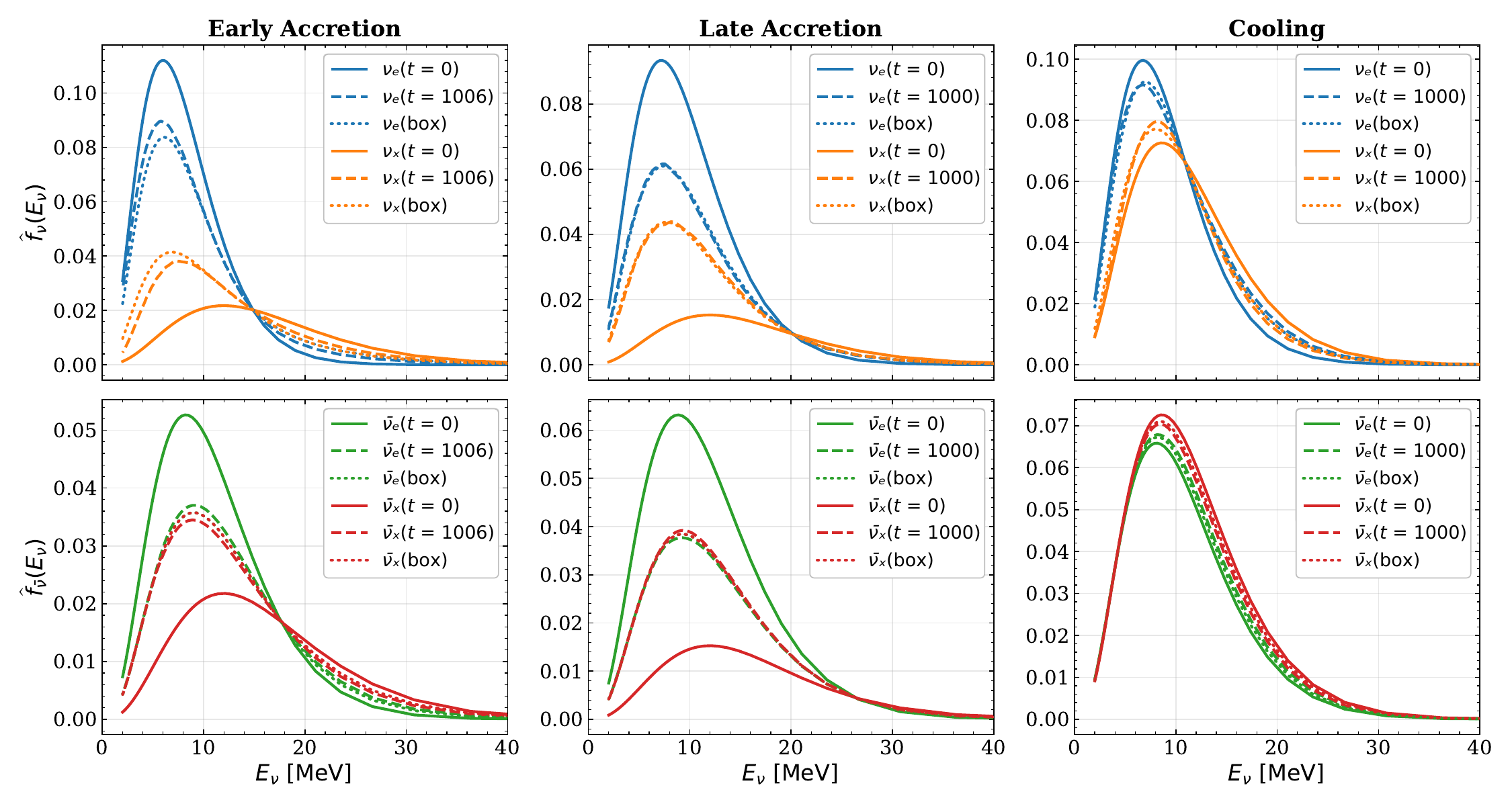}
\caption{Same as Fig.~\ref{fig:energy_spectra_no}, but for IO.}
\label{fig:energy_spectra_io}
\end{figure*}

In Figs.~\ref{fig:w_spec_NO} and~\ref{fig:w_spec_IO}, we compare the spatially averaged and angle-integrated initial and asymptotic lepton-number spectra $G_\omega$ of the three phases in NO and IO, respectively.
These plots show that the final-state $G_\omega$ can generally still contain spectral crossings, clearly exemplified by the early accretion phase in both NO and IO, although the depths of the negative $G_\omega$ are substantially reduced by SFC.
We speculate that this may be consistent with the conclusion of Ref.~\cite{fiorillo2026lepton} that the presence of multiple weak spectral crossings may not guarantee a flavor instability.
For the late accretion and cooling phases, slight flavor overconversion generally occurs for the lobes whose $G_\omega$ are negative initially, consistent with the result of Ref.~\cite{padillagay2025flavor} for single-energy cases. 
Whether this overconversion is related to the initially imposed strong flavor instability associated with deep spectral crossings, see, e.g., Ref.~\cite{wu2021collective,fiorillo2026quasilinear} for fast flavor conversions, which may not actually develop in realistic astrophysical environments, remains to be clarified by future work.  

Among all cases, nearly perfect flavor equipartition only occurs for the two middle lobes associated with small $|\omega|$ (high-energy $\nu$ and $\bar\nu$ above the crossing energy) in the late accretion phase, although for all cases the high-energy neutrino modes (small $\omega$) reach a final state relatively close to equipartition. 
In contrast, the largest positive lobe in the neutrino sector (large $\omega$) experiences the least flavor conversion to preserve the approximate E-XLN conservation condition.
Also noted is that the positive $G_\omega$ lobe in $\omega<0$ (antineutrinos) in both cooling phase cases remains largely positive instead of reaching flavor equipartition\footnote{This is in contradiction with Ref.~\cite{goimilgarcia2026quasisteady}, which claims flavor equipartition for antineutrinos.}.
Overall, these results clearly show that the extent of flavor conversions due to SFC is energy dependent. 
As a consequence, the impact of SFC on physical quantities that are more sensitive to the neutrino spectrum at high energy, such as heating and reaction rates, cannot be well captured by treatments adopting the single-energy assumption.

\section{Impact on energy spectra, heating rates and $Y_e^{\rm eq}$}
\label{sec:astro_implications}

The energy-dependent modification of the neutrino energy spectra discussed in the previous section has several physical consequences. 
In addition to the change of the (anti)neutrino spectra, relevant to the neutrino signals expected from the next galactic supernova, it also leads to the modification of the charged-current absorption rates on free nucleons in the post-shocked region as well as in the neutrino-driven wind.
The spectral change affects both the heating rate behind the supernova shock and the equilibrium electron fraction, and hence potentially impacts the supernova explosion dynamics and the associated wind nucleosynthesis. 
In this section, we compute the coarse-grained, angle-integrated, post-SFC energy spectra for all cases that we have examined, and use them to evaluate the corresponding changes to the neutrino heating rates as well as the neutrino absorption equilibrium $Y_e$.
We note that these evaluations are meant to provide clues about how SFC may affect relevant quantities for CCSN physics within the considered scenarios.
The actual impact on CCSN dynamics and nucleosynthesis requires a self-consistent implementation of SFC into CCSN simulations, which is beyond the scope of this work.

\subsection{Energy spectra}

We use Eq.~\eqref{eq:fhat_from_phi} to reconstruct the physical energy spectra of different species and show the results in Figs.~\ref{fig:energy_spectra_no}
and~\ref{fig:energy_spectra_io} for all three phases in NO and IO, respectively, where the physical energy units have been restored. 
As indicated previously for $G_\omega$, above the crossing energy in the neutrino sector, the final $\nu_e$ and $\nu_x$ spectra become much more similar, approaching flavor equipartition for both NO and IO.
In contrast, below the crossing energy the conversion remains partial due to the E-XLN conservation. 
For antineutrinos, the post-SFC energy spectra are rather close to full equipartition over the entire energy range in both accretion phases, whereas the cooling-phase spectra differ substantially from equipartition for both NO and IO, consistent with Figs.~\ref{fig:w_spec_NO} and ~\ref{fig:w_spec_IO}.
We note that the energy-dependent imprint of SFC on the neutrino energy spectra is distinct from the effect of FFC; see, e.g., Fig.~21 of Ref.~\cite{xiong2024fast}.
As a result, detailed extraction of CCSN neutrino energy spectra of different species, if achievable, can potentially be used as diagnostics for the dominant mode of collective flavor conversions.

\subsection{Heating Rates}
\label{subsec:heating}

The neutrino heating rates due to the charged-current absorption of $\nu_e$ and $\bar\nu_e$ on free nucleons are given by~\cite{xiong2019activesterile}
\begin{equation}\label{eq:qen}
    \dot{q}_{\nu_en}\!\propto \!\int_0^\infty[E_\nu+(\Delta-m_e)](E_\nu+\Delta)^2\!\Big(\! 1-\frac{W_{\nu_e} E_\nu}{m_N} \!\Big)\!\widehat f_{\nu_e}(E_\nu)~dE_\nu ,
\end{equation}
and 
\begin{equation}\label{eq:qep}
    \dot{q}_{\bar{\nu}_ep}\! \propto \!\int_{E_{th}}^\infty[E_\nu-(\Delta-m_e)](E_\nu-\Delta)^2\!\Big(\! 1-\frac{W_{\bar{\nu}_e} E_\nu}{m_N} \!\Big)\!\widehat f_{\bar\nu_e}(E_\nu)dE_\nu, 
\end{equation}
where $\Delta = m_n - m_p$ is the neutron--proton mass difference, $E_{th}=\Delta+m_e$ is the absorption energy threshold, and 
\begin{equation}\label{eq:Wnu}
W_{\nu_e(\bar{\nu}_e)}=\frac{2[1+5g_A^2-(+)2g_A(1+f_2)]}{1+3g_A^2}
\end{equation}
is the weak-magnetism correction~\cite{horowitz1999charge} with $g_A=1.27$ and $f_2=3.71$. 
To evaluate the change of these heating rates due to SFC, we compute them using the initial and final spectra shown in Figs.~\ref{fig:energy_spectra_no} and \ref{fig:energy_spectra_io} and denoted the corresponding rates by $\dot q^i$ and $\dot q^f$, respectively.

\begin{table}[t]
\centering
\caption{Ratios of the final (post-SFC) to initial heating rates $\dot{q}^f/\dot{q}^i$ for the three CCSN phases in normal and inverted mass orderings. The values of the ratios for different mass orderings are separated by slashes.}
\label{tab:qratio}
\begin{tabular}{lcc}
\hline\hline
Phase & $\dot{q}_{\nu_e n}^f/\dot{q}_{\nu_e n}^i$ (NO/IO) & $\dot{q}_{\bar\nu_e p}^f/\dot{q}_{\bar\nu_e p}^i$ (NO/IO) \\
\hline
Early accretion & $1.84\,/\,1.62$ & $1.39\,/\,1.31$ \\
Late accretion  & $1.12\,/\,1.10$ & $0.89\,/\,0.90$ \\
Cooling         & $1.47\,/\,1.53$ & $1.12\,/\,1.18$ \\
\hline\hline
\end{tabular}
\end{table}

Table~\ref{tab:qratio} shows $\dot{q}^f_{{\nu}_en}/\dot{q}^i_{{\nu}_en}$ and 
$\dot{q}^f_{\bar{\nu}_ep}/\dot{q}^i_{\bar{\nu}_ep}$ for all three phases and both mass orderings. 
The \nue~heating rate is enhanced in all phases, with the largest increase during early accretion: $84\%$ in NO and $62\%$ in IO. 
This enhancement is driven by near flavor equipartition at high energies above the crossing energy, which raises the high-energy tail of the post-conversion $\widehat f_{\nu_e}$. 
The \nuebar~heating rate is enhanced by $39\%$ ($31\%$) during early accretion and by $12\%$ ($18\%$) during cooling in NO (IO). 
For late accretion, it instead decreases by $11\%$ in NO and $10\%$ in IO, because the high-energy tails of the $\bar\nu_e$ and $\bar\nu_x$ spectra are too similar. 
Consequently, the reduction of the lower-energy part of the $\bar\nu_e$ spectrum results in a net decrease of the $\bar\nu_e$ heating rate.

\subsection{Absorption equilibrium electron fraction}
\label{subsec:ye}

The neutrino absorption equilibrium electron fraction, which is a key quantity for nucleosynthesis conditions in neutrino-driven winds~\cite{qian1996nucleosynthesis}, is defined by 
\begin{equation}\label{eq:ye}
    Y_e^{\rm eq}=\Big(1+\frac{\lambda_{\bar{\nu}_ep}}{\lambda_{{\nu}_en}}\Big)^{-1}\ ,
\end{equation}
with
\begin{align}\label{eq:lam}
    \lambda_{{\nu}_en} &\propto\int_0^\infty(E_\nu+\Delta)^2\Big( 1-\frac{W_{\nu_e} E_\nu}{m_N} \Big)\widehat f_{\nu_e}(E_\nu)~dE_\nu\ ,\\
    \lambda_{\bar{\nu}_ep} &\propto\int_{E_{th}}^\infty(E_\nu-\Delta)^2\Big( 1-\frac{W_{\bar\nu_e} E_\nu}{m_N} \Big)\widehat f_{\bar\nu_e}(E_\nu)~dE_\nu\ ,
\end{align}
denoting the neutrino absorption rates on nucleons.
Table~\ref{tab:ye} lists its value computed using the initial and final spectra for all three phases. 

\begin{table}[t]
\centering
\caption{Initial and final (post-SFC) equilibrium electron fractions for the three CCSN phases.}
\label{tab:ye}
\begin{tabular}{lccc}
\hline\hline
Phase & $Y_e^{\rm eq,i}$ & $Y_e^{\rm eq,f}$ (NO) & $Y_e^{\rm eq,f}$ (IO) \\
\hline
Early accretion & $0.597$ & $0.624$ & $0.614$ \\
Late accretion  & $0.580$ & $0.615$ & $0.609$ \\
Cooling         & $0.607$ & $0.643$ & $0.641$ \\
\hline\hline
\end{tabular}
\end{table}

Here, we see that SFC leads to an increase in $Y_e^{\rm eq}$ for both NO and IO in all phases by $0.017$--$0.036$.  
This is due to the reduced value of $\lambda_{\bar{\nu}_ep}/\lambda_{{\nu}_en}$ as $\lambda_{{\nu}_en}$ is enhanced more than $\lambda_{\bar{\nu}_ep}$.
As a result, SFC can lead to more proton-rich conditions in the $\nu$-driven outflows and potentially favor the production of proton-rich nuclei, particularly if the conditions for the $\nu p$-process~\cite{frohlich2006neutrinoinduced,pruet2006nucleosynthesis, Friedland:2025lge, Friedland:2026asd} can be realized in the outflow.

\section{A Phenomenological recipe for quasistationary energy spectra}
\label{sec:recipe}

An important question is whether the coarse-grained quasistationary state due to SFC can be predicted by knowing the initial spectrum of $G(\omega,v_z)$ without performing a full nonlinear QKE calculation. 
While we find that the final $\langle P_3(\omega,v_z)\rangle_z$ shows a rather complicated structure, which may be related to the shape of the eigenfunctions of the initial unstable modes for some cases (see Appendix~\ref{app:eigenfit}), it appears challenging to find a simple analytical fit to approximate the full $G(\omega,v_z)$ satisfactorily. 
In this section, we instead investigate whether the coarse-grained, angle-averaged spectra can be well approximated by simple analytical prescriptions. 

Assuming that SFC tends to erase (or largely reduce) the initial spectral crossings as shown in earlier sections, we attempt to approximate the final shape of $G_\omega$ by imposing a box-like, $\omega$-dependent flavor survival probability, motivated by fast flavor conversion studies~\cite{xiong2023evaluating,zaizen2023simple}\footnote{This recipe differs from that proposed in Ref.~\cite{goimilgarcia2026quasisteady} for cases without E-XLN angular crossings, where full flavor equipartition for antineutrinos is assumed.}. 
We define $I_-$ and $I_+$ as the absolute area of the negative and positive portions of the lepton-number spectrum, respectively, so that
\begin{equation}\label{eq:AB}
I_- \equiv \Big|\!\!\int_{G_\omega < 0}\! d\omega\,G_\omega\Big|\ ,\qquad I_+ \equiv \int_{G_\omega > 0}\! d\omega\,G_\omega\ . 
\end{equation}
We then formulate the asymptotic survival probability by 
\begin{equation}\label{eq:Pomega}
P^{\rm box}(\omega) = \begin{cases}
p_\mathrm{eq}\ , & G_\omega > 0\ ,\\[3pt]
1 - (1 - p_\mathrm{eq})\,\dfrac{I_+}{I_-}\ , & G_\omega < 0\ ,
\end{cases}\quad (I_- \geq I_+)\ ,
\end{equation}
and 
\begin{equation}
P^{\rm box}(\omega) = \begin{cases}
1 - (1 - p_\mathrm{eq})\,\dfrac{I_-}{I_+}\ , & G_\omega > 0\ ,\\[3pt]
p_\mathrm{eq}\ , & G_\omega < 0\ ,
\end{cases}\quad (I_+ > I_-)\ ,
\end{equation}
where $p_{\rm eq}=1/2$ enforces flavor equipartition for $G_\omega>0$ ($G_\omega<0$) when $I_-\geq I_+$ ($I_-<I_+$).
The associated spectrum is then
\begin{equation}\label{eq:Gfinal}
G_\omega^\mathrm{box} = \bigl(2P^{\rm box}(\omega) - 1\bigr)\,G_\omega^\mathrm{initial}\ .
\end{equation}

We show $G_\omega^\mathrm{box}$ as well as the corresponding $\widehat f_\nu(E_\nu)$ together with the nonlinear results in Figs.~\ref{fig:w_spec_NO}--\ref{fig:energy_spectra_io}. 
Since the prescription depends only on the initial $G_\omega$, the box prediction is identical for NO and IO. Figs.~\ref{fig:w_spec_NO} and~\ref{fig:w_spec_IO} show that the resulting spectrum qualitatively agrees with the simulation outcome, although it fails to capture the flavor overconversion in the late accretion and cooling phases, as well as the residual negative $G_\omega$ in the early accretion phase. 
Similarly, the predicted energy spectra displayed in Figs.~\ref{fig:energy_spectra_no} and~\ref{fig:energy_spectra_io} show generally good agreement with the simulation results. 

To quantify the performance of the above box recipe, we define $\dot{q}_\mathrm{est}/\dot{q}_\mathrm{sim}$ as the ratio of the heating rate obtained from the box-generated spectrum to that obtained from the simulation result, and report the values for all cases in Table~\ref{tab:recipe_perf}.
The table shows that the box recipe accurately reproduces the changes in the heating rates within $10\%$ for most cases. 
The only exceptions are $\dot q_{\nu_e}$ for IO and $\dot q_{\bar\nu_e}$ for NO in the early accretion phase, where $G_\omega^{\rm box}$ deviates the most from the simulated $G_\omega$ at small $|\omega|$ (see Figs.~\ref{fig:w_spec_NO} and \ref{fig:w_spec_IO}).

\begin{table}[t]
\centering
\caption{Ratio of the box-estimated to simulated heating rate, $\dot{q}_\mathrm{est}/\dot{q}_\mathrm{sim}$. 
Each entry lists NO/IO, and values close to 1 indicate good agreement.}
\label{tab:recipe_perf}
\begin{tabular}{lcc}
\hline\hline
Phase & $\nu_e$ (NO/IO) & $\bar\nu_e$ (NO/IO) \\
\hline
Early accretion & $1.067/1.214$ & $0.859/0.910$ \\
Late accretion  & $1.009/1.031$ & $0.988/0.984$ \\
Cooling         & $0.979/0.942$ & $0.975/0.928$ \\
\hline\hline
\end{tabular}
\end{table}

In addition, we further quantify the agreement of the energy spectra in Figs.~\ref{fig:energy_spectra_no} and~\ref{fig:energy_spectra_io} by defining the simulation-normalized net percentage differences 
\begin{equation}
\begin{aligned}
\Delta_{\alpha}^{\rm box}
&\equiv
100\times\frac{\displaystyle\int dE_\nu\,
\left|\widehat f_\alpha^{\rm box}(E_\nu)-\widehat f_\alpha^{\rm sim}(E_\nu)\right|}
{\displaystyle\int dE_\nu\,
\widehat f_\alpha^{\rm sim}(E_\nu)},\\
\Delta_{\rm all}^{\rm box}
&\equiv
100\times\frac{\displaystyle\sum_\alpha\int dE_\nu\,
\left|\widehat f_\alpha^{\rm box}(E_\nu)-\widehat f_\alpha^{\rm sim}(E_\nu)\right|}
{\displaystyle\sum_\alpha\int dE_\nu\,
\widehat f_\alpha^{\rm sim}(E_\nu)},
\end{aligned}
\label{eq:box_spectrum_l1}
\end{equation}
where the summation in the second equation runs over all species.

Table~\ref{tab:box_spectrum_l1} reports $\Delta_{\alpha}^{\rm box}$ and $\Delta_{\rm all}^{\rm box}$ for all cases. 
Consistently, the early accretion phase shows the largest deviation, particularly for $\nu_x$.  
Otherwise, the agreement is generally within $5$--$10\%$.

\begin{table}[H]
\centering
\caption{Simulation-normalized percentage differences between the box-predicted and simulated energy spectra [Eq.~\eqref{eq:box_spectrum_l1}]. 
The paired columns list the electron/heavy-flavor values.}
\label{tab:box_spectrum_l1}
\setlength{\tabcolsep}{3pt}
\begin{tabular}{llccc}
\hline\hline
Phase & MO & $\Delta_{\nu_e}^{\rm box}/\Delta_{\nu_x}^{\rm box}$ & $\Delta_{\bar\nu_e}^{\rm box}/\Delta_{\bar\nu_x}^{\rm box}$ & $\Delta_{\rm all}^{\rm box}$ \\
\hline
Early accretion & NO & $7.36/13.20$ & $7.02/7.74$ & $8.55$ \\
                & IO & $7.21/12.13$ & $3.35/3.41$ & $6.61$ \\
Late accretion  & NO & $3.06/3.79$  & $4.17/3.95$ & $3.70$ \\
                & IO & $1.95/2.47$  & $1.26/1.23$ & $1.74$ \\
Cooling         & NO & $2.93/3.32$  & $0.80/0.73$ & $2.00$ \\
                & IO & $3.61/4.29$  & $2.61/2.50$ & $3.28$ \\
\hline\hline
\end{tabular}
\end{table}

\section{Discussion and Conclusions}
\label{sec:conclusion}

In this work, we have extended our previous single-energy study of SFC to a multi-energy, multi-angle system for three neutrino spectra representing different CCSN phases. 
The linear stability analysis and nonlinear QKE calculations show that SFC is predominantly triggered by spatially inhomogeneous instabilities. 
For the two accretion phase cases, the maximum growth rates are substantially larger in IO than in NO, while the ordering dependence is weaker for the cooling phase case. 

We find that the resulting coarse-grained, quasistationary flavor conversion probability depends on energy, angle, and the neutrino mass ordering. 
In contrast to the single-energy system, the initial spectral crossings are not completely erased in the coarse-grained, angle-averaged final state. 
High-energy modes at small $|\omega|$ generally undergo stronger flavor conversion, while lower-energy modes retain a larger fraction of their initial flavor, independent of the mass ordering. 
The final spectra can exhibit residual crossings as well as flavor overconversion in some cases. 
These features demonstrate that the post-SFC spectra cannot be represented by a simplified monochromatic treatment. 

We have also investigated the impact of SFC on the charged-current neutrino interaction rates. 
Across all cases considered in this work, the $\nu_e$ absorption heating rates are enhanced by $10\%$--$84\%$. 
For $\bar\nu_e$, the heating rate exhibits a weaker enhancement or even a slight suppression, since its energy spectrum is closer to that of $\bar\nu_x$. 
For both mass orderings and all three phases, SFC increases $Y_e^{\rm eq}$ by $0.017$--$0.036$. 
This indicates that SFC can potentially affect neutrino heating as well as result in more proton-rich conditions relevant to the yields of proton-rich nuclei in the neutrino-driven wind, if the spectral changes observed in this work are representative. 

We have also tested a phenomenological box prescription that estimates the asymptotic coarse-grained and angle-averaged energy spectra directly from the initial spectra. 
Although the prescription is independent of the mass ordering and cannot reproduce the flavor overconversion and residual crossings obtained in simulations, it reasonably approximates the simulated energy spectra, with relative spectral differences for individual species and heating-rate deviations generally $\lesssim \mathcal{O}(5$--$20\%)$.

While our work has improved the understanding of the evolution and the local coarse-grained quasistationary state of SFC for supernova neutrinos, several aspects remain to be further investigated. 
The formal stability condition for SFC with multiple spectral crossings remains to be established. 
Our initial conditions and the ratio of $\bar\omega_{\nu_\alpha}/\mu$ are chosen such that the system starts with strong flavor instabilities in the ``non-resonant'' regime~\cite{fiorillo2024theory}.  
While a recent work has demonstrated ``bipolar-like'' evolution of SFC under weak instabilities in the ``resonant'' regime~\cite{fiorillo2026singlewave}, it is important to conduct self-consistent global QKE simulations including collisions for SFC that take into account the natural appearance of weak instabilities~\cite{johns2025subgrid,fiorillo2024fast,xiong2026neutrino} as well as their development in a spatially varying matter background~\cite{bhattacharyya2025role,zaizen2026fast,fiorillo2026flavomons}, to fully assess the evolution and impact of SFC in astrophysical environments.  
With these in mind, our work represents an important step toward the goal of extending the subgrid models developed for FFC~\cite{nagakura2024bhatnagargrosskrook,xiong2025robust} to account for the impact of SFC on global neutrino transport.

\textit{\textbf{Acknowledgments.}}---
IPG would like to thank Julien Froustey for insightful discussions. 
HHC and MRW acknowledge support of the National Science and Technology Council, Taiwan under Grant Nos. 111-2628-M-001-003-MY4, 115-2112-M-001-054-MY5, and Academia Sinica under Project No.~AS-IV-114-M04. 
MRW also acknowledges support of the Physics Division of the National Center for Theoretical Sciences, Taiwan.
IPG is supported by NSF Physics Frontier Center Award number 2020275. 
SA was supported by the German Research Foundation (DFG) through the Collaborative Research Centre ``Neutrinos and Dark Matter in Astro- and Particle Physics (NDM),'' Grant No.\ SFB-1258\,--\,283604770, and under Germany’s Excellence Strategy through the Cluster of Excellence ORIGINS EXC-2094-390783311.
ZX acknowledges support of the European Research Council (ERC) through grant NeuTrAE (No.~101165138). The work is partially funded by the European Union. Views and opinions expressed are however those of the authors only and do not necessarily reflect those of the European Union or the European Research Council Executive Agency. Neither the European Union nor the granting authority can be held responsible for them.
IPG, SA, and ZX would like to thank Academia Sinica for their hospitality during the workshop ``Collective Neutrino Oscillations in Supernovae and Neutron Star Mergers'' in Taipei, where this project was partly developed. 
This work used ASGC (Academia Sinica Grid- computing Center) Distributed Cloud resources, which is supported by Academia Sinica. 
This work made use of the following software packages: \texttt{Jupyter}~\citep{perez2007ipython,kluyver2016jupyter}, \texttt{matplotlib}~\cite{hunter2007matplotlib}, \texttt{numpy}~\cite{harris2020array}, \texttt{python}~\cite{vanrossum2009python}, \texttt{scipy}~\cite{virtanen2020scipy,gommers2024scipy}, and \texttt{Mathematica}~\cite{wolfram2024mathematica}.
MRW acknowledges the use of OpenAI's Codex in this study to assist with modifying plotting scripts and to cross-check numerical values by post-processing the simulation results. 

\bibliography{references_collective_oscillations}

\appendix

\section{Numerical convergence and resolution}
\label{appendix:resolution}

\begin{figure}[ht]
\centering
\includegraphics[width= 0.9\columnwidth]{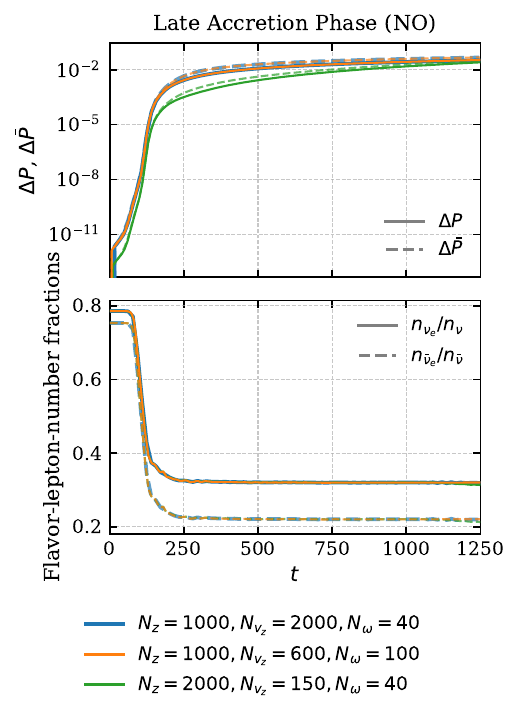}
\caption{Evolution of $\Delta P(t)$ and $\Delta \bar P(t)$ (upper panel; Eq.~\eqref{eq:polarization_norm_errors}) as well as $n_{\nu_e}/n_\nu$ and $n_{\bar\nu_e}/n_{\bar\nu}$ for late accretion in NO using different values of $(N_z,N_{v_z},N_\omega)=(1000,2000,80)$, $(1000,600,200)$, and $(2000,150,80)$, respectively.
$t$ is in units of $\mu^{-1}$.
}
    \label{fig:DeltaP_evolution}
\end{figure}

We have examined the numerical convergence of our results with different choices of $N_z$, $N_\omega$, and $N_{v_z}$, and show in Fig.~\ref{fig:DeltaP_evolution} comparisons of the deviations of the polarization-vector norms from unity in the neutrino and antineutrino sectors (upper panel), defined by
\begin{align}
 \Delta P(t)&\equiv\left\langle\left|\,|\mathbf P(t,z,\omega,v_z)|-1\right|\right\rangle_{z,v_z,\omega>0},\nonumber\\
 \Delta\bar P(t)&\equiv\left\langle\left|\,|\mathbf P(t,z,\omega,v_z)|-1\right|\right\rangle_{z,v_z,\omega<0}.
 \label{eq:polarization_norm_errors}
\end{align}
We also show the evolution of $n_{\nu_e}/n_\nu$ and $n_{\bar\nu_e}/n_{\bar\nu}$ (lower panel) for $(N_z, N_{v_z}, N_\omega)=(1000, 2000, 80)$, $(1000, 600, 200)$, and $(2000, 150, 80)$.
These comparisons clearly show no visible differences in $n_{\nu_e}/n_\nu$ and $n_{\bar\nu_e}/n_{\bar\nu}$, and $\Delta P(t)$ and $\Delta \bar P(t)$ remain $\lesssim\mathcal{O}(10^{-2})$ until the end of the simulation.

\begin{figure*}[ht]
\centering
\includegraphics[width= 0.9\textwidth]{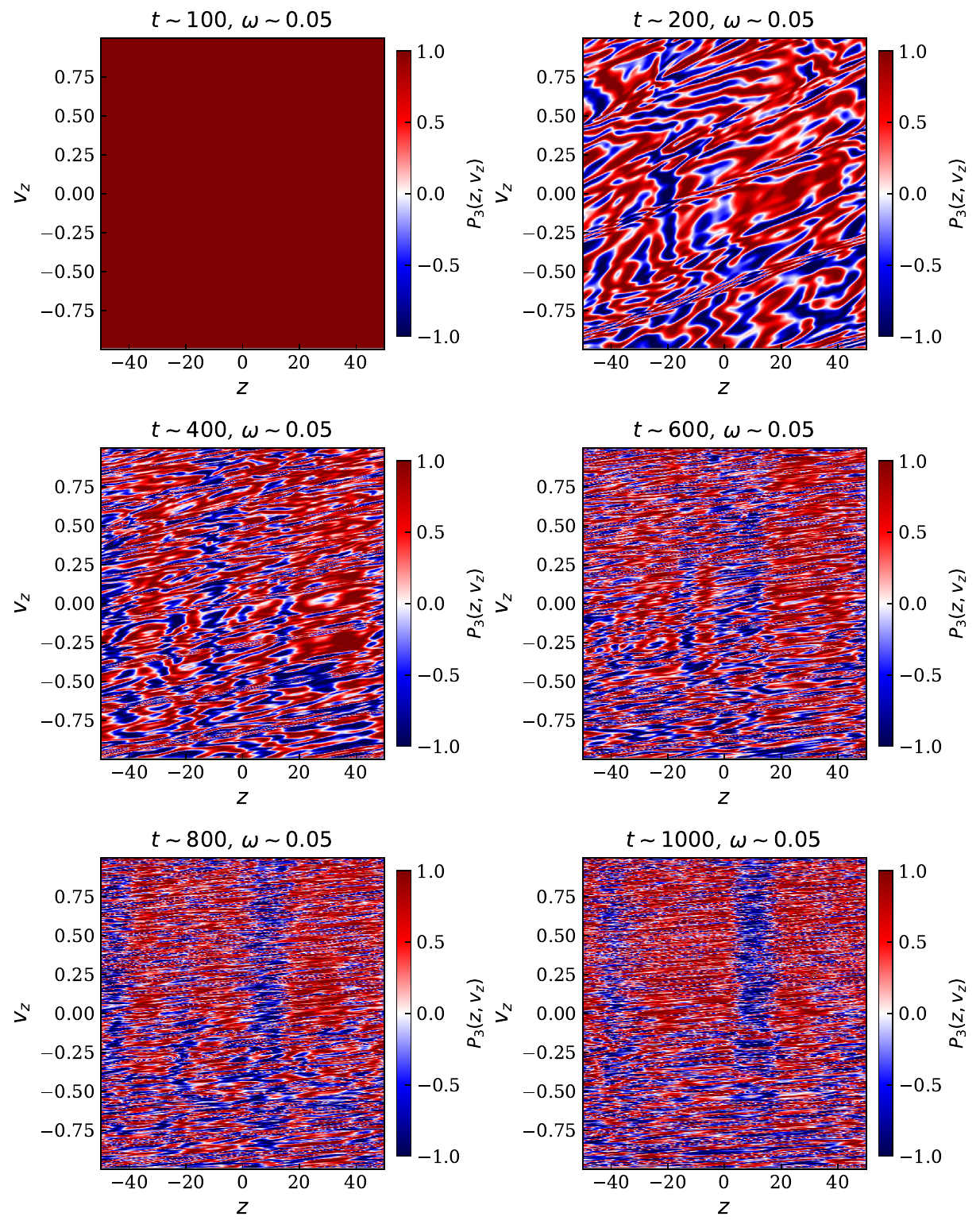}
\caption{Snapshots of $P_3(z,v_z)$ for $\omega\simeq0.05$ for late accretion in IO with $(N_z,N_{v_z},N_\omega)=(1000,600,200)$ at $t\simeq 100,200,400,600,800$, and $1000$.
The quantities $t$ and $\omega$ are in units of $\mu^{-1}$ and $\mu$, respectively.}
\label{fig:vz_cascade}
\end{figure*}

While the averaged quantities show nearly perfect agreement for all cases, we caution that following the continuous late-time evolution requires a large $N_{v_z}$ after the system seems to have reached a coarse-grained quasistationary state. 
This is illustrated by Fig.~\ref{fig:vz_cascade}, which shows $P_3(v_z,z)$ for $\omega=0.05$ at different times for the late accretion case in IO using $(N_z,N_{v_z},N_\omega)=(1000,600,200)$. 
For this case, the ratios $n_{\nu_e}/n_\nu$ and $n_{\bar\nu_e}/n_{\bar\nu}$ have already come very close to their quasistationary values at $t\sim 200$ (see Fig.~\ref{fig:time_evolution}), and they continue decreasing slowly before reaching final values at $t\gtrsim 600$. 
For evolution at late times, Fig.~\ref{fig:vz_cascade} shows that the coherent pattern developed at $t\sim 200$ gets sheared into narrower and narrower filaments, particularly in the $v_z$ direction; fully resolving this cascade therefore requires a large $N_{v_z}$.  

We note that the behavior of the $v_z$ cascade appears to be case dependent.
For instance, the cooling cases and the early accretion case in NO retain a more coherent large-scale structure throughout the entire evolution than the late accretion case in IO discussed above.
Interestingly, the cases that suffer from severe $v_z$ cascades seem to be the ones whose $\langle P_3(\omega,v_z) \rangle$ lacks correlation with $W(\omega,v_z)$ discussed in Sec.~\ref{app:eigenfit}. 
Whether these phenomena are linked by an underlying mechanism requires further investigation beyond the scope of this work.

\begin{figure*}[ht]
\centering
\includegraphics[width= 1.0\textwidth]{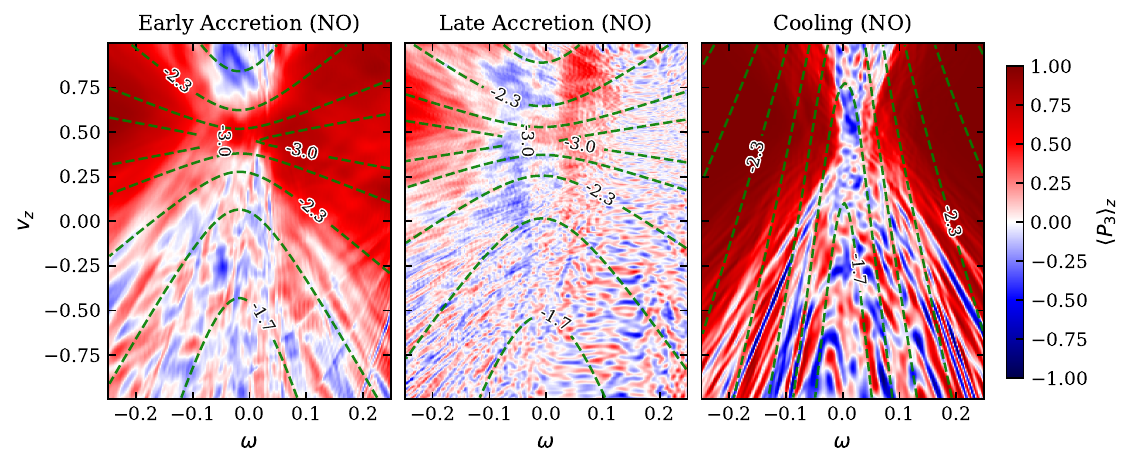}
\includegraphics[width= 1.0\textwidth]{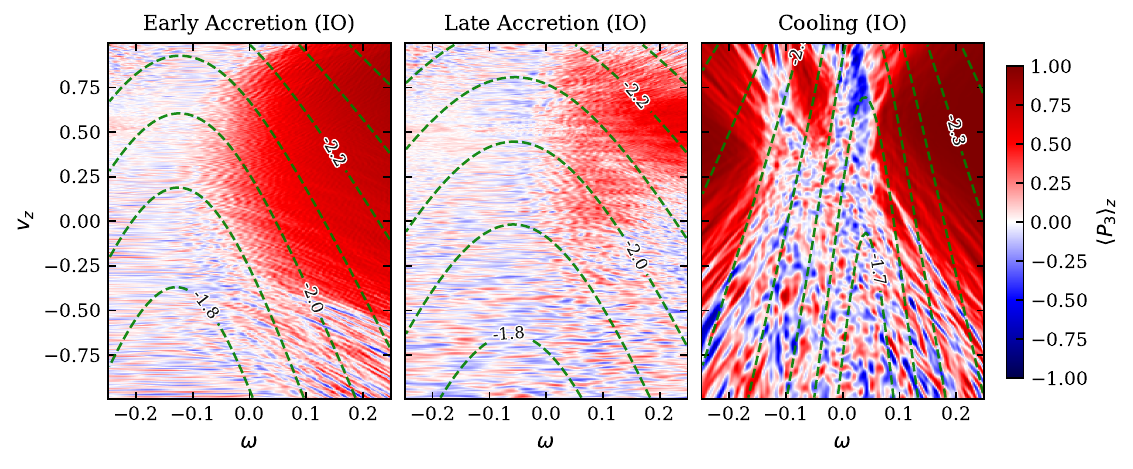}
\caption{Spatially averaged polarization vector component $\langle P_3\rangle_z$ in the $(\omega,v_z)$ plane (color) for early accretion (left panels), late accretion (middle panels), and cooling (right panels). The upper and lower rows show NO and IO, respectively, at the final times used in Figs.~\ref{fig:energy_spectra_no} and~\ref{fig:energy_spectra_io}.
Also shown are the contours of $\log_{10}W(\omega, v_z)$ (green dashed) computed using Eq.~\eqref{eq:Wfit}; see text for details.
The quantities $t$ and $\omega$ are in units of $\mu^{-1}$ and $\mu$, respectively.}
\label{fig:P3_W_contour}
\end{figure*}

\section{Coarse-grained asymptotic phase-space distributions}
\label{app:eigenfit}

We show in Fig.~\ref{fig:P3_W_contour} the final coarse-grained function $\langle P_3(\omega,v_z) \rangle_z$ for all six cases (three phases and two mass orderings). 
These panels show that while the coarse-grained, angle-integrated energy spectra appear to be rather similar in NO and IO (see Figs.~\ref{fig:w_spec_NO} and ~\ref{fig:w_spec_IO}), the underlying full phase-space distributions can be very different. 

We have also investigated whether the shape of the unstable eigenmodes leaves any imprints on the final distribution of $\langle P_3(\omega,v_z) \rangle_z$.  
The green-dashed curves in Fig.~\ref{fig:P3_W_contour} show the contours of a function $W(\omega, v_z)$, which represents the growth-rate-weighted superposition of the 15 fastest-growing modes among all $k$ obtained in Sec.~\ref{sec:lsa}:

\begin{equation}\label{eq:Wfit}
    W(\omega, v_z) \equiv  \frac{\sum_{j=1}^{15} e^{\gamma_j t_W}\,|u_j(\omega, v_z)|}{\sum_{j=1}^{15} e^{\gamma_j t_W}},
\end{equation}
where $\gamma_j=\mathrm{Im}\,\Omega_j$ and $u_j(\omega,v_z)$ are, respectively, the growth rate and normalized eigenvector of mode $j$. 
For the reference time $t_W$, we set $t_W=20\gamma_{\rm max}^{-1}$ using $\gamma_{\rm max}^{-1}$ listed in Table \ref{tab:gamma_max} so that it represents the time when the system has nearly reached the quasistationary state.  
The numbers attached to the green curves shown in the figure are $\log_{10}W$.

Comparing the shapes of $W$ to $\langle P_3 \rangle$ suggests that regions experiencing less flavor conversions (redder) generally have more negative values of $\log_{10}W$.
In particular, $W$ qualitatively captures the dominant shape of $\langle P_3 \rangle$ for the early accretion case in NO, a part of the phase space in the late accretion case in NO (upper left corner), as well as the two cooling phase cases.  
However, they do not seem to show any correlations for the two accretion phases in IO.
Thus, while $W$ identifies where the instability initially develops, the nonlinear mode--mode coupling may have prevented it from predicting the detailed quasistationary pattern.

\end{document}